\documentclass[aps, prb,twocolumn]{revtex4}
\usepackage{color}   % colors
\usepackage{graphicx}
\usepackage{amsmath}
\usepackage{amssymb} %required for \Box
\usepackage{epstopdf}
\begin{document}

\title{Transport Conductivity of Graphene at RF and Microwave Frequencies}
\author{S. A. Awan$^{1,2}$, A. Lombardo$^1$, A. Colli$^3$, G. Privitera$^1$, T. Kulmala$^1$, J. M. Kivioja$^3$, M. Koshino$^4$, A. C. Ferrari$^1$}
\email{acf26@eng.cam.ac.uk}
\affiliation{$^1$ Cambridge Graphene Centre, University of Cambridge,Cambridge CB3 0FA, UK}
\affiliation{$^2$School of Computing, Electronics and Mathematics, Plymouth University, Plymouth, PL4 8AA, UK}
\affiliation{$^3$ Nokia Technologies, Broers Building, Cambridge, CB3 0FA, UK}
\affiliation{$^4$ Department of Physics, Tohoku University, Sendai, 980-8578, Japan}

\begin{abstract}
We measure graphene coplanar waveguides from direct current (DC) to $f$=13.5GHz and show that the apparent resistance (in the presence of parasitic impedances) has an $\omega^2$ dependence (where $\omega=2\pi f$), but the intrinsic conductivity (without the influence of parasitic impedances) is frequency-independent. Consequently, in our devices the real part of the complex alternating current conductivity is the same as the DC value and the imaginary part$\sim0$. The graphene channel is modeled as a parallel resistive-capacitive network with a frequency dependence identical to that of the Drude conductivity with momentum relaxation time$\sim2.1$ps, highlighting the influence of alternating current (AC) electron transport on the electromagnetic properties of graphene. This can lead to optimized design of high-speed analogue field-effect transistors, mixers, frequency doublers, low-noise amplifiers and radiation detectors.
\end{abstract}
\maketitle
\section{\label{Intro}Introduction}
\begin{figure*}
\centerline{\includegraphics[width=180mm]{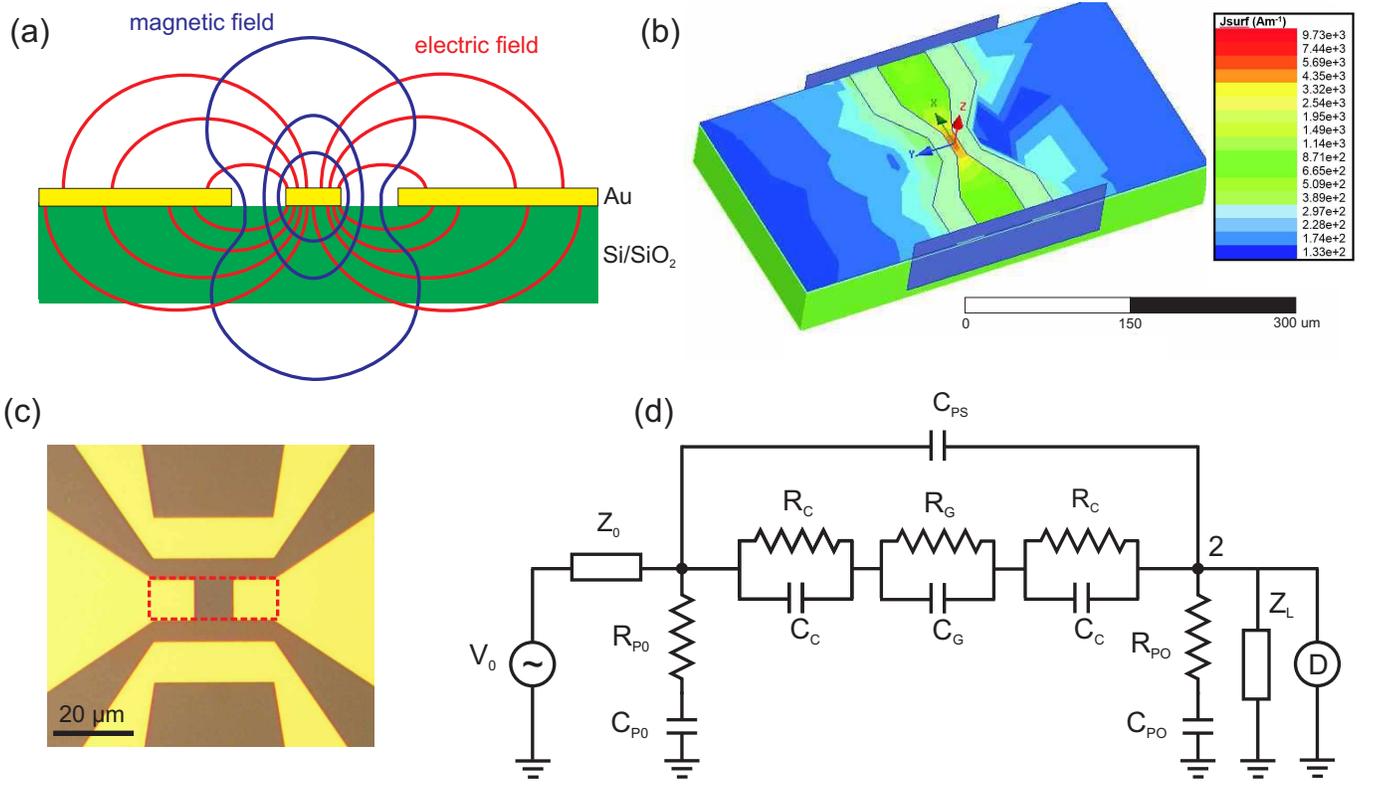}}
\caption{(a) Distribution of electric and magnetic fields in a AuCPW with TEM mode propagation, (b) HFSS simulation of surface current density (in units of $Am^{-1}$) distribution in a AuCPW at 13.5GHz, with 1$V_{pk}$ excitation voltage between the ground and signal conductors, (c) GCPW in the central signal conductor (red dashed line) and (d) its equivalent lumped-parameter model of the GCPW with source and detector illustrating measurement of the $S_{21}$ scattering parameter.}
 \label{graph:CPW}
\end{figure*}
Graphene is a promising material for high-frequency electronic applications, ranging from DC to THz\cite{NovoS306,FerrNANO7,VicaNM11,SpirAPL104,Sordan}, such as transistors\cite{WuN472,WuNL12,LiaoN467,LinNL9,LinS327,PallAPL99,MoonIEDL30,LemmIEDL28}, low-noise amplifiers\cite{DasIEEEP}, mixers\cite{WangIEDL31}, frequency doublers\cite{WangIEDL31,YangACSN4} and microwave radiation detectors\cite{DragJAP107}. This is because of its high carrier mobility ($>100,000$ $cm^2V^{-1}s^{-1}$ at room temperature\cite{Mayo11}), ambipolar transport\cite{NovoS306}, high Fermi velocity $v_F=1.1\times10^{-6}$m/s [\onlinecite{NovoN438}], current carrying capacity\cite{YuNL12} ($\sim$1.8$\times10^{9}Acm^{-2}$) and thermal conductivity\cite{BalaNL8} ($\sim$5000W$m^{-1}K^{-1}$). Power dissipation through a channel of resistance R carrying radio-frequency (1-300MHz) and microwave signals (0.3-300GHz) is also an important parameter, particularly for applications in high-speed electronics, such as transistors and low noise amplifiers. Any increase of channel resistance at higher frequencies (above the DC value), $\Delta R(\omega)=[R_{ac}(\omega)-R_{dc}]/R_{dc}$, will contribute excess noise and impact the signal-to-noise ratio at both the component level and when integrated into a complete system-on-chip\cite{Bertz} or microwave monolithic integrated circuit (MMICs)\cite{DealMM9}. The dependence of $\Delta R$ on frequency therefore needs to be determined accurately to ensure that the power dissipation does not become prohibitive for applications in a given frequency range (such as microwaves or THz). Power dissipation, e.g., degrades the signal-to-noise ratio in high-frequency detectors\cite{DragJAP107}. Metals and superconductors show strong frequency dependence of their surface impedance which manifests as electromagnetic losses$\cite{DuffBOOK,AwanIPSMT149}\propto\omega^{2}$ and skin effect losses\cite{DuffBOOK}$\propto\omega^{\frac{1}{2}}$, leading to an additional AC resistance.

The AC (or dynamic) conductivity $\sigma(\omega)$ of single layer graphene (SLG) from DC to optical frequencies can be modeled by the Kubo formalism\cite{Kubo12,Ando71,GusyCM19,HansJAP103,KoshSSC149,KoshPRB87} as[\onlinecite{FaloEPJB56,KoppeNL11}]:
\begin{eqnarray}\nonumber
\label{Kubo}
\sigma (\omega)&=&\frac{2e^2T}{\pi \hbar}\frac{i}{\omega+i\tau^{-1}}log\left[2cosh\left(\frac{E_F}{2k_BT}\right)\right]\\&& +\frac{e^2}{4\hbar}\left[H(\omega/2)+\frac{4i\omega}{\pi} \int_0^\infty\frac{H(\epsilon)-H(\omega/2)}{\omega^2 - 4 \epsilon^2}d\epsilon\right]\nonumber\\
\end{eqnarray}
where T is the temperature, $\hbar$ the reduced Planck's constant, \textit{i} the imaginary unit, $E_F$ the Fermi energy, $k_B$ the Boltzmann's constant, and $H(\epsilon)$ is[\onlinecite{FaloEPJB56,KoppeNL11}]:
\begin{eqnarray}
\label{H}
H(\epsilon)=\frac{sinh\left(\frac{\hbar \epsilon}{k_BT}\right)}{cosh\left(\frac{E_F}{k_BT}\right)+cosh\left(\frac{\hbar \epsilon}{k_BT}\right)}
\end{eqnarray}
Eq.\ref{Kubo} consists of intra- and inter-band contributions, corresponding to the first and second term respectively. The conductivity depends on the energy of the incident RF radiation, such that the interband term corresponds to electron-hole (e-h) pair generation and recombination events, whereas the intraband converges to the Drude model for T=0K. In the DC to 13.5GHz range, relevant for devices and applications such as transistors, mixers and low noise amplifiers, the inter-band transitions are negligible and Eq.\ref{Kubo} can be rewritten as:
\begin{eqnarray}\nonumber
\label{intra}
\sigma_{intra}(\omega, \mu_c,\gamma,T)&=&\frac{ie^2k_BT}{\pi\hbar^2(\omega+i\gamma)}\left[\frac{\mu_c}{k_BT}+\right. \\&& \left. 2\ln\left(e^{-\mu_c/k_BT}+1\right)\right]
\end{eqnarray}

where $\gamma=\tau^{-1}$ is the electron scattering rate (in units of $s^{-1}$)  due to electron interactions with impurities, defects, phonons and disorder and $\mu_c$ is the chemical potential. Eq. (\ref{intra}) can be rewritten in the Drude form (at room temperature and constant $\mu_c$ and $\gamma$) as [\onlinecite{HornPRB83}]:
\begin{equation}
\label{drude}
\sigma (\omega) = \sigma _1(\omega) - i \sigma _2 (\omega) = (i W_D) (\pi \omega + i \pi \gamma)^{-1}
\end{equation}
where $\sigma _1 (\omega)$ and $\sigma _2 (\omega)$ are the real and imaginary components of the conductivity. The prefactor $W_D$, known as the Drude weight\cite{HornPRB83,AbedPRB84}, is:
\begin{eqnarray}\nonumber
\label{drude_weight}
W_D(\mu_c, T) &=& (e^2 k_B T/ \hbar^2) [ \mu_c/(k_B T)\\&& + 2 ln [ 1 + e^{-\mu_c/(k_B T)}]
\end{eqnarray}
$W_D$ for graphene is different from conventional metals due to its linear energy-wavevector dispersion (in contrast with $W_D =\pi n e^2/m^*$ in metals, where $m^*$ is the carrier effective mass and its number density\cite{KitteBOOK}). Eq.(\ref{drude}) suggests that, for $\omega\ll2\pi / \tau$, graphene's conductivity should be frequency-independent and approximately equal to the DC conductivity $\sigma _0$. Given that $\gamma$ of SLG is in the order of 1 - 20 THz (depending on doping, material quality, i.e. exfoliated/chemical vapor deposited (CVD)), which corresponds to $\tau\sim0.05-1$ps [\onlinecite{HornPRB83,TanPRL99,MittNP11, LinQE20}], graphene's conductivity should be frequency-independent up to$\sim$0.5-1THz. However, experimental confirmation of this frequency-independent response of graphene to transport RF and microwave signals has not been reported, to the best of our knowledge.

We integrate SLG into coplanar waveguide (CPW) transmission lines in order to investigate its RF and microwave transport properties. The CPW transmission lines (Fig.\ref{graph:CPW} a,b) consist of a central signal conductor in close proximity with two ground conductors\cite{WenIEEETM17}. These are ideal for investigating the RF to mm-wave electromagnetic transport properties of a variety of materials and devices\cite{Poza}, since their properties are well established theoretically\cite{Poza, WenIEEETM17} and experimentally\cite{WenIEEETM17}. Compared to microstrips\cite{Poza}, CPWs enable quasi-transverse electromagnetic (TEM) wave propagation (where quasi-TEM refers to the presence of small but finite longitudinal electric and magnetic field components), low dispersion of its characteristic impedance, low cross-talk (or interference with any nearby devices) and broadband (DC to mm-wave\cite{Poza}) operation. Furthermore, CPWs can also be used as a building block for the integration of passive and active components into complete systems-on-chip\cite{Bertz} or MMICs\cite{DealMM9} (which integrate a range of functionalities, such as mixing, amplification, switching etc. at microwave frequencies). CPWs allow accurate measurement of RF properties, since their electromagnetic properties are traceable to well-established and extensively used Short-Open-Load-Thru (SOLT) reference standards (linked to the benchmark National Institute for Standards and Technology multiline Thru-Reflect-Line procedures\cite{MarkITMTT39}), provided by instrument manufactures and national metrology institutes to enable calibration of vector network analyzers (VNAs)\cite{MarkITMTT39}.

Several groups have integrated graphene with CPWs in order to investigate its electromagnetic transport properties and reported a change in the DC to AC conductivity. Ref.[\onlinecite{MoonNJP12}] reported $\Delta \sigma (\omega)\sim-11\%$ from DC to 10 GHz, whereas in Ref.[\onlinecite{LeeAPL100}] $\Delta \sigma (\omega)\sim-6.4\%$ from DC to 13.5 GHz and in Ref.[\onlinecite{SkulAPL99}] $\Delta \sigma (\omega)\sim-1.9\%$ from DC to 13.5 GHz. These results contrast the predictions of Eqs.\ref{Kubo}-\ref{drude_weight}.

Here, we report the design, fabrication and characterization of graphene CPWs (GCPWs) up to 13.5GHz. We extract their transmission line parameters and compare them with Au waveguides without the presence of graphene. We measure an intrinsic resistance and intraband conductivity of graphene frequency-independent up to 13.5GHz. This contrasts the frequency-dependent resistance of metals and superconductors at RF and microwave frequencies and can be used to design and develop broadband RF devices based on graphene.
\section{\label{Results}Results and discussion}
We first design, characterize and optimize a set of Au CPWs. The width of the CPWs is fixed at$\sim400\mu m$ having a pitch of the Ground-Signal-Ground (GSG) contact pads$\sim$150$\mu m$ to match the GSG probe tips connected to semi-rigid coaxial cables interfaced to an Agilent N5230C VNA, with an upper frequency limit of 13.5GHz. The CPWs length is optimized to$\sim500\mu m$ in order to accommodate the tapering of the signal conductor from its contact pad of dimensions 100$\mu m$ x 100$\mu m$ to a 10$\mu m$ x 10$\mu m$ SLG sample. The contacts are formed by evaporating 2nm Cr/80 nm Au. The optimized Au CPWs show excellent broadband RF and microwave transmission properties from DC to 13.5GHz, as discussed later.

\begin{figure}[h!]
\centerline{\includegraphics[width=90mm]{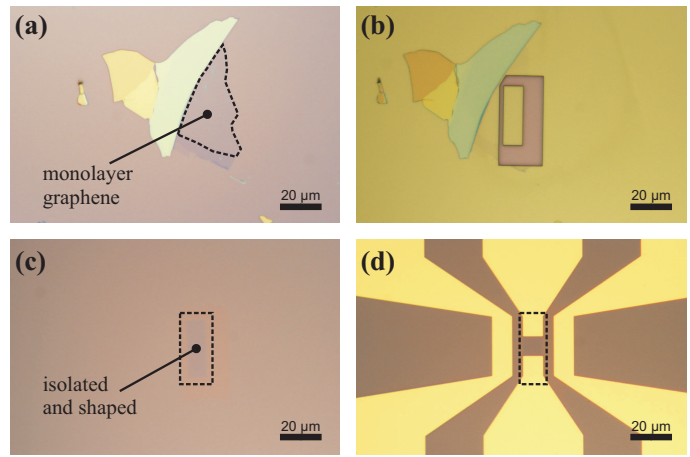}}
\caption{Fabrication of GCPWs: (a) Exfoliated graphite flakes on SiO$_2$ typically comprise a mixture of mono and multilayer graphene. (b) The area to be used as channel is defined by e-beam lithography and plasma etching. (c) The polymer mask is detached, resulting in the removal of unwanted thick flakes, leaving a shape-defined and isolated SLG. (d) Cr/Au contacts are then fabricated to form the GCPW.}
\label{graph:Inverse_transfer}
\end{figure}

We then fabricate GCPWs identical to the optimized AuCPWs except for the signal conductor having a gap for positioning SLG. SLG flakes are prepared by micro-mechanical cleavage of graphite\cite{BonaMT15, NovoPNAS} on a high resistivity ($>10k\Omega cm$) Si+ 285nm SiO$_2$ substrate. The single layer nature of the flakes is confirmed by a combination of optical microscopy\cite{CasiNL7} and Raman spectroscopy\cite{FerrPRL97,FerrNN8}. Polymethyl methacrylate (PMMA) is then spin-coated onto the substrate. A frame with the desired shape is subsequently defined via e-beam lithography. After resist development, a mild oxygen plasma is used to remove the uncovered SLG parts. This results in an island of desired rectangular shape (30$\mu m$x10$\mu m$) isolated from the rest of the polymer film. The latter is then removed by immersion in de-ionized water, while the lithographically defined island remains on the substrate. PMMA is then dissolved leaving an isolated SLG flake\cite{BonaMT15}. Cr/Au contacts are then deposited, as for Fig.\ref{graph:Inverse_transfer}.
\begin{figure}[h!]
\centerline{\includegraphics[width=90mm]{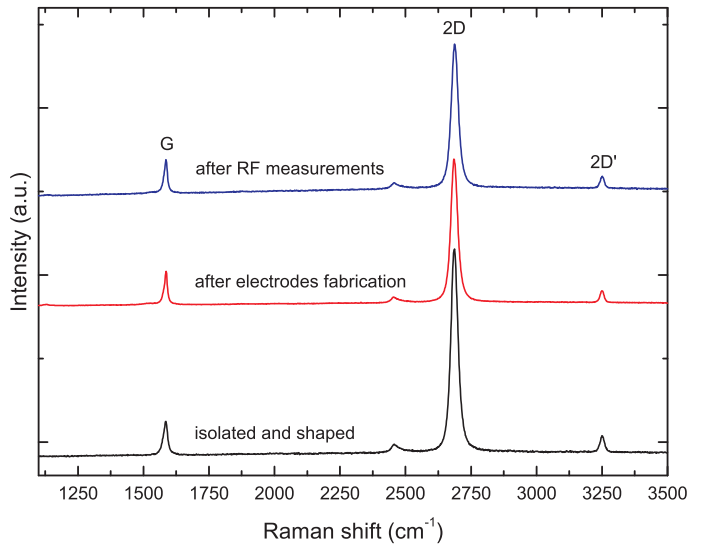}}
\caption{Raman spectra of (a) SLG flake prior to shaping. (b) SLG flake after shaping. (c) The same flake after fabrication of the contacts and (d) after the RF measurements.}
\label{graph:Raman}
\end{figure}

Raman spectroscopy is used at every stage of the device fabrication process and after RF measurements. Fig.\ref{graph:Raman} plots representative Raman spectra (acquired in the same spot) of shaped SLG, the same SLG after contact fabrication and post RF measurements. By analyzing the position of the  G peak, Pos(G), its full width at half maximum, FWHM(G), the position of the 2D peak, Pos(2D), as well as  the intensity and area ratios I(2D)/I(G) and A(2D)/A(G), it is possible to conclude that the sample is p-doped\cite{PisaNM6, BaskPRB80, DasNN3}, with a carrier concentration$\sim2-3\times 10^{12}$cm$^{-2}$ [\onlinecite{BaskPRB80,DasNN3}] and a Fermi energy$\sim200$meV  [\onlinecite{BaskPRB80,DasNN3}]. After contact fabrication, only small changes of Pos(G) and Pos(2D) ($\sim$1cm$^{-1}$) occur, while I(2D)/I(G) is reduced from 6 to 4, indicating that that the doping of the sample increases during the process, but still remains below$\sim$5$\times 10^{12}cm^{-2}$. Doping remains unchanged after RF measurements, as indicated by no changes in peak positions, FWHMs, intensity and area ratios\cite{PisaNM6,DasNN3,CasiAPL91}. No significant D peak is detected at any stage of the fabrication process nor after the RF measurements, proving the high structural quality of the flakes and the non-invasiveness of the measurements.

The GCPWs are first characterized at DC using a parameter analyzer in a 2-probe configuration since SLG is positioned in the signal conductor (i.e. source-drain configuration). In order to contact the SLG, the CPWs are tapered from the contact pads to match the width of the selected SLG flakes, whilst maintaining the 50$\Omega$ characteristic impedance of the waveguide, as shown in Fig\ref{graph:CPW}(c). The corresponding equivalent lumped-parameter (or discrete electrical components) model for the GCPW, together with the parasitic impedances, is shown in Fig.\ref{graph:CPW}(d). The lumped-parameters are based on the physical layout and geometry and are equivalent to a transmission line model\cite{AwanBOOK,DuffBOOK}. The electrical properties of a CPW are related to its impedance \textit{Z}, which depends on the geometry and dielectric properties of the surrounding medium (such as Si substrate, SiO$_2$ layer and air). The CPW impedance is given by\cite{WenIEEETM17} $Z = 60\pi/\sqrt{\epsilon_{re}}(\psi(\chi)+\psi'(\chi))$ where $\psi$ and $\psi'$ are elliptic integrals of the first kind and their complement, $\epsilon_{re}$ is the relative permittivity of the substrate and $\chi=b / (b +2d)$, where $b$ is the width of the central conductor and $d$ is the width of the gap between the central and ground conductors. In Fig.\ref{graph:CPW}(d), the capacitance C$_C$ and resistance R$_C$ account for the two contacts (i.e. source and drain) on the SLG, while the SLG channel is modeled by a parallel resistance and capacitance, R$_G$ and C$_G$. C$_{PS}$ accounts for the capacitive coupling between the two leads contacting the SLG channel. R$_{P0}$ and C$_{P0}$ denote the coupling between the SLG contacts and the outer ground electrodes. Fig.\ref{graph:CPW}(a) shows the schematic electric and magnetic field distributions when an even-mode (or transverse electromagnetic mode) is excited in the waveguide. Parasitic odd-modes (or non-TEM modes) can also be excited if, e.g., the two ground conductors are at different potentials as a result of improper or non-planar GSG contact. Fig.\ref{graph:CPW}(b) plots a finite element simulation (using the High Frequency Structures Simulation (HFSS) software with 1$V_{pk}$ excitation voltage between the ground and signal conductors at 13.5 GHz) of a tapered AuCPW at 13.5GHz optimized (through simulation of the waveguides with varying geometrical designs) to enable integration of SLG as the channel material. The figure shows the distribution of the surface current density at 13.5GHz in the waveguide. HFSS solves Maxwell's equations at each mesh in the simulation domain to calculate the current density distribution on metallic and dielectric surfaces\cite{Poza}.

\begin{figure}[h!]
\centerline{\includegraphics[width=90mm]{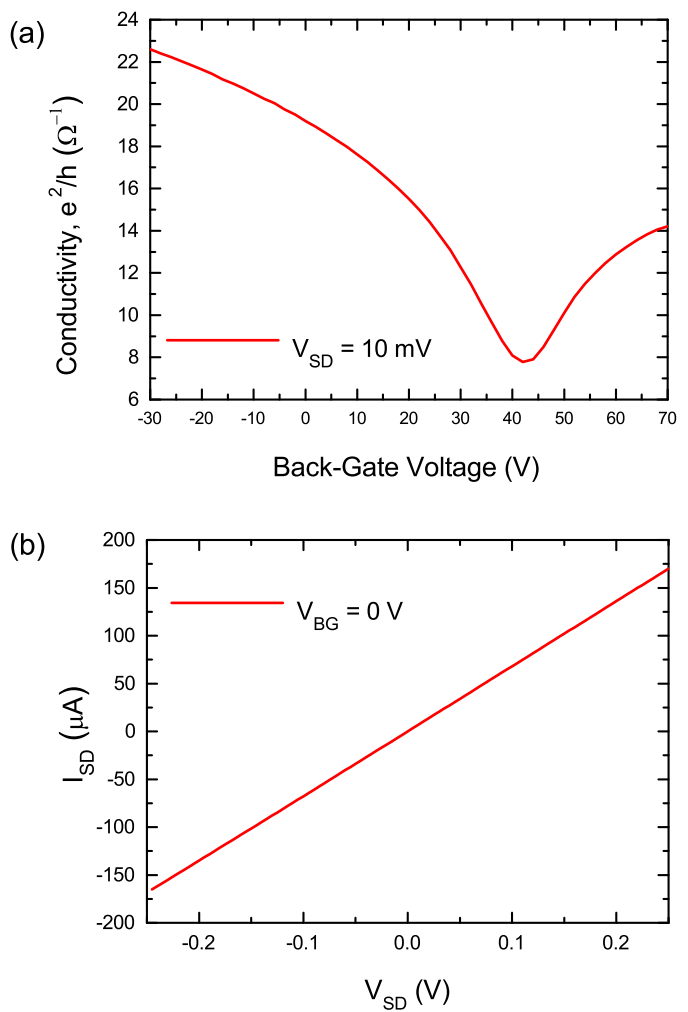}}
\caption{(a) Two-probe DC conductivity of the SLG channel in the CPW at room temperature, (b) linear $I_{SD}$ versus $V_{SD}$, with $R_{2P}\sim1.47 k\Omega$}
\label{graph:Conductivity}
\end{figure}

Fig.\ref{graph:Conductivity}(a) plots the two-probe DC conductivity and back-gate voltage dependence measured on our GCPW between the input and output signal conductors (or source and drain electrodes). The SLG channel dimensions are length $a=10\mu m$ and width $b=10\mu m$, and the overlap under the Cr/Au contacts is 10$\mu m$ in length and width (precisely defined through e-beam lithography). The SLG channel in the waveguide is p-doped ($\sim$200$meV$), consistent with the Raman analysis. The Dirac-point is detected at $V_D\sim40V$ and corresponds to a minimum conductivity $\sigma_{0}^{min}\sim8e^{2}/h$ and a sheet resistance $R_{S}\sim3.3 k\Omega /\Box$. Fig.\ref{graph:Conductivity}(b) indicates a linear dependence of $I_{SD}$ on $V_{SD}$, with two-probe resistance $R_{2P}\sim1.47 k\Omega$ at zero back-gate potential. All RF and microwave measurements are performed with a fixed zero back-gate potential. The corresponding carrier density and mobility are estimated as $n=\alpha(V_{BG}-V_D)$ with $\alpha=7.2 \times 10^{10}cm^{-2}$ [\onlinecite{NovoS306}], $n\sim2.9 \times 10^{12}cm^{-2}$ and $\mu\sim2200 cm^2V^{-1}s^{-1}$, respectively, and the Fermi level $E_{F}=\hbar v_F \sqrt{\pi |n|}$$\sim$200$meV$. The carrier mean-free path can be estimated from\cite{TanPRL99} $l=(\hbar /e)\mu\sqrt{\pi |n|}$ as $\sim44$nm using the measured carrier concentration at DC and room temperature, indicating diffusive transport (comparable to the $\sim 40-100nm$ typically found in SLG devices on SiO$_{2}$ at room temperature\cite{AdamPNAS47,TanPRL99}).
\begin{figure*}
\centerline{\includegraphics[width=180mm]{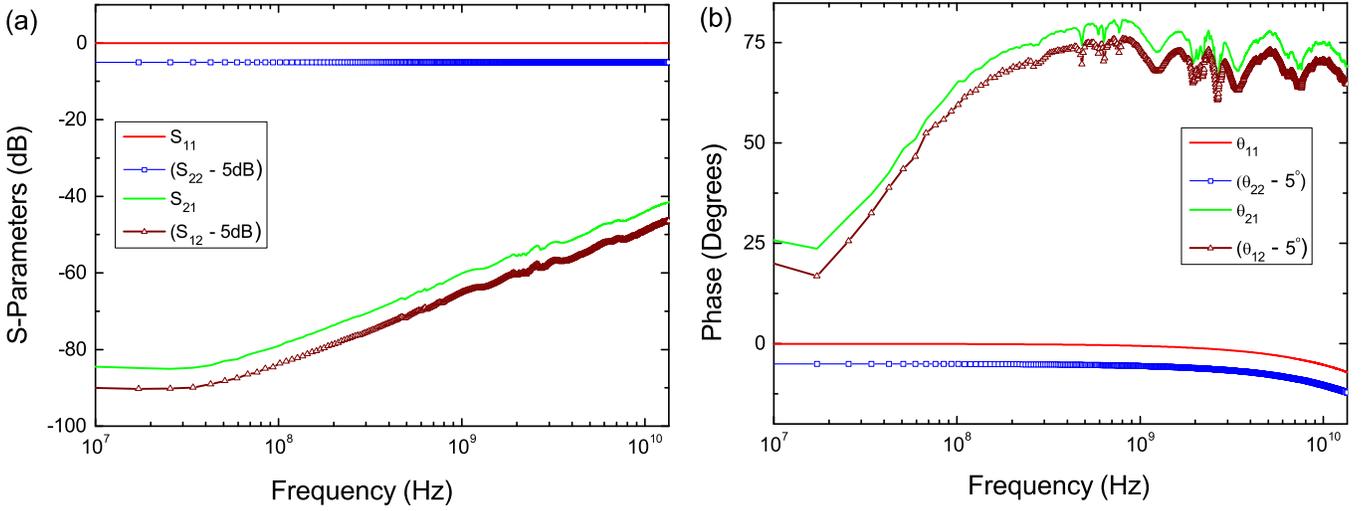}}
\caption{Measured scattering parameters (a) magnitude and (b) phase for our Open CPW at frequencies up to 13.5GHz. The magnitude and phase data are offset by 5dB and 5$^{\circ}$, respectively, for clarity.}
\label{graph:open}
\end{figure*}

Subsequent to the DC measurements, our GCPWs are characterized at RF and microwave frequencies by measuring the magnitude and phase of the scattering-parameters, $S_{jk} \angle \Theta_{jk}$, in a two-port configuration, where $(j,k)\in (1,2)$, with $[\textbf{S}]$ a 2x2 S-parameters matrix. For a two-port network with characteristic impedance \textit{Z}$_{0}$ the impedance matrix $[\textbf{Z}]$ in terms of S-parameter matrix $[\textbf{S}]$ and the identity matrix \textbf{[I]} is given by\cite{Poza}:
\begin{equation}
\label{eq:zed_matrix}
[\textbf{Z}] = Z_0(\textbf{[I]}+[\textbf{S}])(\textbf{[I]}-[\textbf{S}])^{-1}
\end{equation}
The waveguide S-parameters are measured using a VNA calibrated using SOLT (Short, Open, Load, Thru) standards$\cite {Jarg,Nish,Impa}$ on impedance standard substrates at frequencies up to 13.5GHz. After the VNA calibration, a set of de-embedding devices (Open, Short and Thru) which exclude the SLG channel (but are otherwise identical to the GCPWs) are characterized using the same VNA parameter settings (such as power, intermediate frequency bandwidth, averaging factor, sweep time etc.) as during SOLT measurements. The de-embedding devices enable removal of the effects of parasitic impedances from the apparent (or as-measured) response of the GCPWs\cite{Damb,Cho,Lee}. The Open and Short de-embedding structures are used to extract $C_{PS}$, $R_{PO}$ and $C_{PO}$, as for Fig.\ref{graph:CPW}(d). The Thru de-embedding device is used to evaluate repeatability and consistency of measurements and also enables a comparison with the GCPW results.

For any two-port, passive and linear device, exposed to alternating voltages and currents, an admittance matrix may be defined as\cite{Poza}:
\begin{equation}
\left[ \begin{array}{c} I_1 \\ I_2 \end{array} \right] = \begin{bmatrix} Y_{11} & Y_{12} \\ Y_{21} & Y_{22} \end{bmatrix} \times \left[ \begin{array}{c} V_1 \\ V_2 \end{array} \right]
\label{admittance}
\end{equation}
where $V_{1,2}$ and $I_{1,2}$ are the voltages and currents at nodes (1,2), respectively, Fig.\ref{graph:CPW}(d). The measured scattering parameters enable us to derive the corresponding admittance parameters\cite{LiaoN467} $Y_{jk}^m$, with $(j,k) \in (1,2)$:
\begin{figure*}
\centerline{\includegraphics[width=180mm]{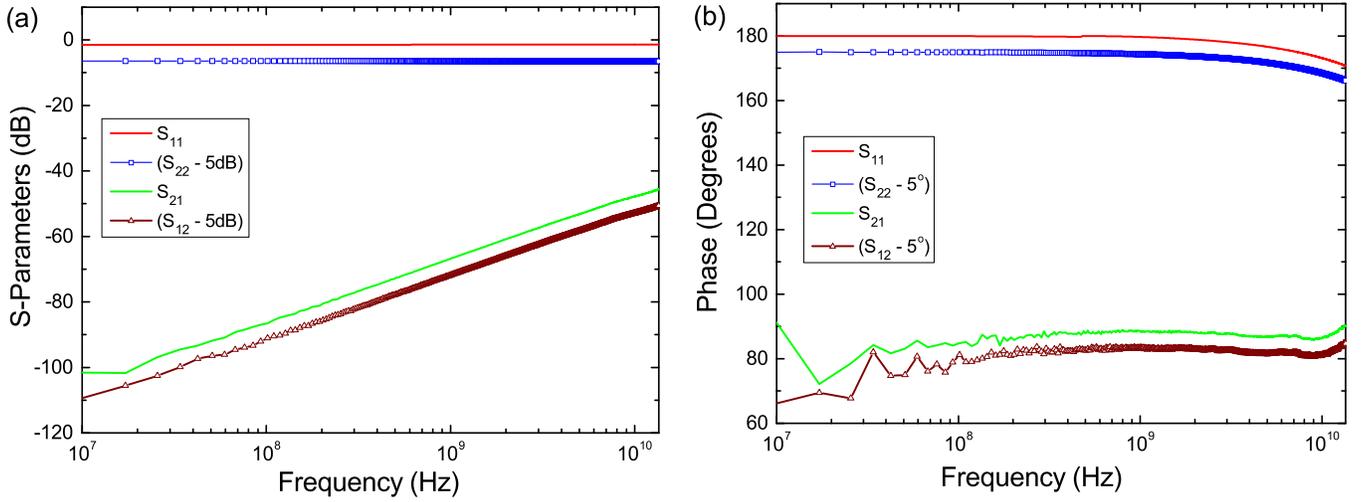}}
\caption{Measured scattering parameters (a) magnitude and (b) phase of the Short CPW at frequencies up to 13.5GHz. The magnitude and phase data are offset by 5dB and 5$^{\circ}$, respectively, for clarity.}
\label{graph:short}
\end{figure*}
\begin{equation}
Y_{jk}^m= Y_0 \begin{bmatrix} \frac{(1-S_{11})(1+S_{22})+S_{12}S_{21}}{\Delta S} & -\frac{2S_{12}}{\Delta S} \\ -\frac{2S_{21}}{\Delta S} &  \frac{(1+S_{11})(1-S_{22})+S_{12}S_{21}}{\Delta S} \end{bmatrix}
\label{parameters}
\end{equation}
\begin{figure*}
\centerline{\includegraphics[width=185mm]{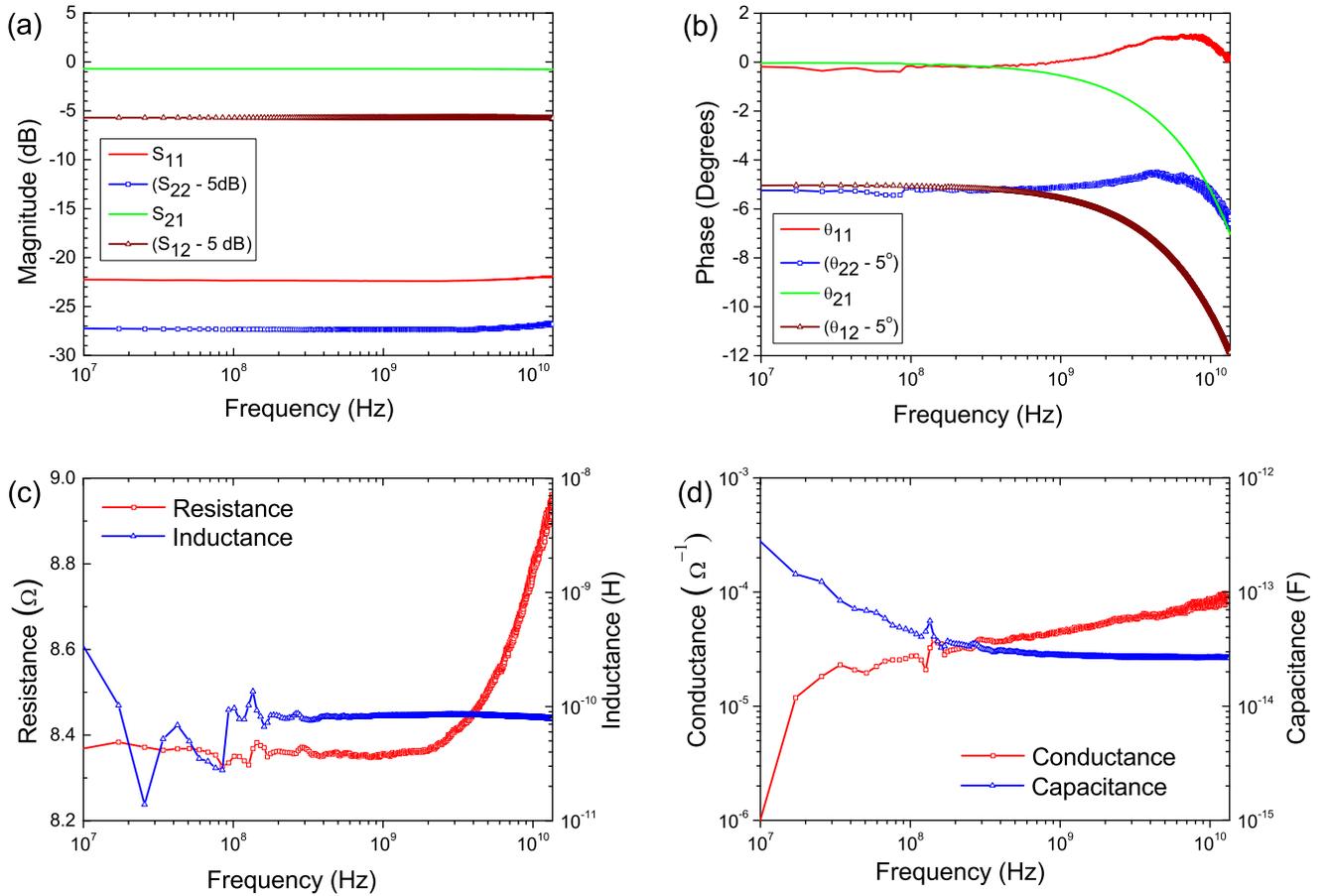}}
\caption{Transport RF and microwave measurements of the scattering parameters (a) magnitude and (b) phase at 0dBm (offset by 5dB and 5$^\circ$, for clarity). The corresponding extracted transmission line parameters for a Au Thru CPW at frequencies up to 13.5GHz for resistance and inductance are shown in (c), whereas the capacitance and conductance are plotted in (d).}
\label{graph:Transport_1}
\end{figure*}
where $Y_0$ is the characteristic admittance and $\Delta S=[(1+S_{11})(1+S_{22})-S_{12}S_{21}]$. The intrinsic admittance matrix $[Y_{jk}^m]$ of our GCPW is extracted by de-embedding the measured impedance matrix of the Open  $[Y_{open}]$ and Short $[Y_{short}]$ devices using\cite{Damb,Cho}:
\begin{equation}
[Y_{jk}] = [(Y_{jk}^m-Y_{open})^{-1}+(Y_{short}-Y_{open})^{-1}]^{-1}
\label{Y}
\end{equation}
\begin{figure*}
\centerline{\includegraphics[width=200mm]{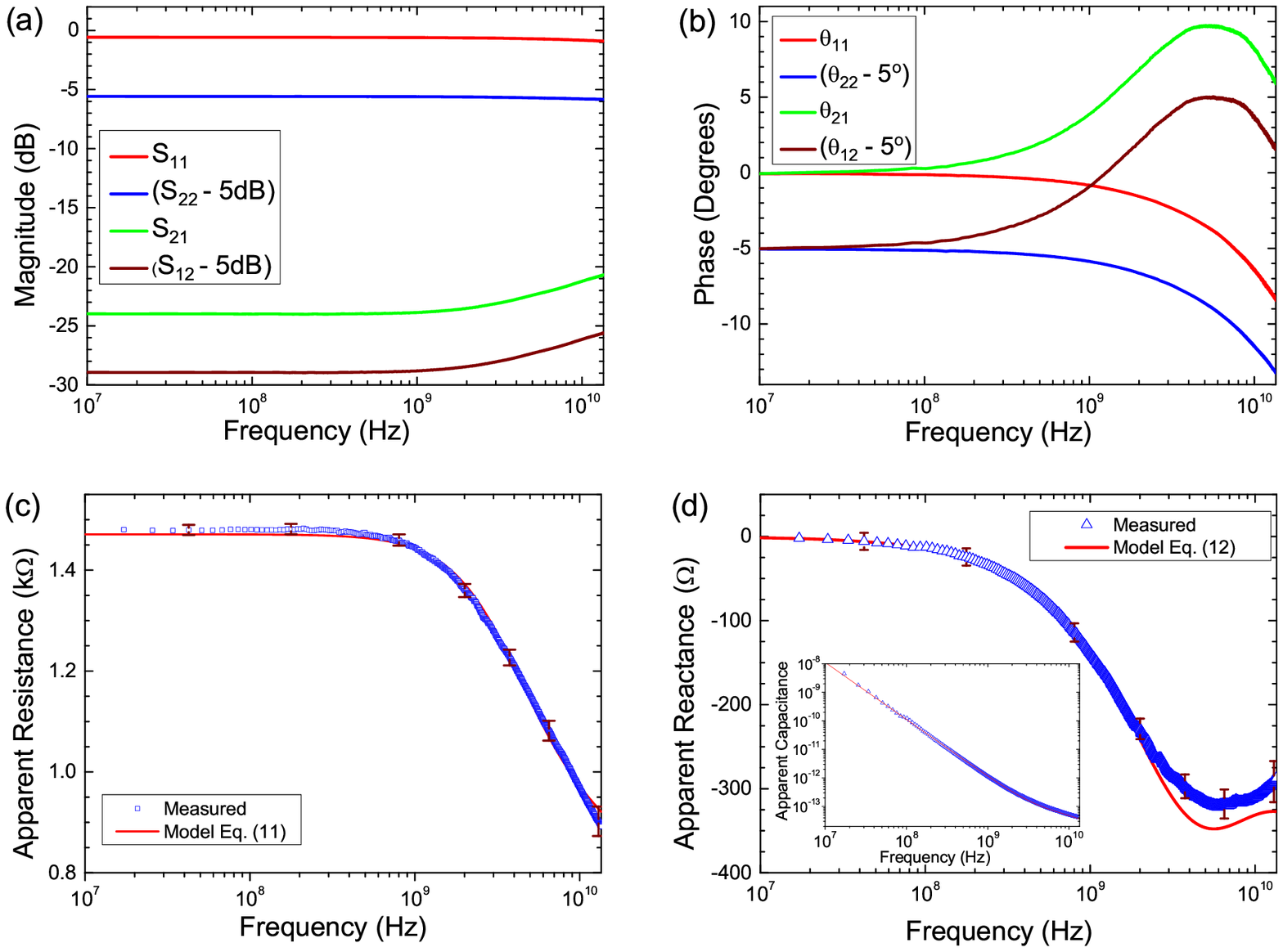}}
\caption{RF and microwave measurements of the scattering parameters (a) magnitude and (b) phase at frequencies up to 13.5GHz with 0dBm incident power, $V_{BG}=0V$ and room temperature (offset by 5dB and 5$^{\circ}$, respectively, for clarity). GCPW (c) apparent resistance and (d) reactance, with solid lines given by Eqs.(\ref{eq:RA}, \ref{eq:XA}). The inset in (d) compares the measured GCPW apparent capacitance with the model of Eq.(\ref{eq:XA}) up to 13.5GHz.}
\label{graph:Transport_2}
\end{figure*}
Fig.\ref{graph:open} plots typical measured S-parameters magnitude and phase for an Open CPW device, and Fig.\ref{graph:short} the corresponding ones for a Short CPW up to 13.5GHz. Using these data in Eq.(\ref{parameters}), the admittance matrices of the Open and Short devices are determined and then inserted into Eq.(\ref{Y}) to derive the de-embedded intrinsic admittance matrix $[Y_{jk}]$. The corresponding impedance matrix is determined using $[Z_{jk}]=[Y_{jk}]^{-1}$ and the component $Z_{21}$ of the intrinsic impedance of the GCPW is:
\begin{equation}
\label{eq:zeta}
Z_{21}=\frac{2R_C}{(1+i\omega C_C R_C)}+\frac{R_{G}}{(1+i\omega C_G R_G)}
\end{equation}
where $R_C$ and $C_C$ are the contact resistance and capacitance between SLG and the Cr/Au contacts and $R_G$ and $C_G$ are the SLG channel resistance and capacitance. The corresponding real and imaginary components of the impedance in Eq.(\ref{eq:zeta}) are given by:
\begin{equation}
\label{eq:RA}
R_A=\frac{2R_C}{(1+\omega^2 C_C^2 R_C^2)}+\frac{R_{G}}{(1+\omega^2 C_G^2 R_G^2)}
\end{equation}
and
\begin{equation}
\label{eq:XA}
X_A=- \bigg[\frac{2\omega C_C R_C^2}{(1+\omega^2 C_C^2 R_C^2)}+\frac{\omega C_G R_G^2}{(1+\omega^2 C_G^2 R_G^2)}\bigg]
\end{equation}
where $R_A$ and $X_A$ are the apparent resistance and reactance of the GCPW, respectively. The second part of Eqs.(\ref{eq:RA}, \ref{eq:XA}) resembles the Drude model even though the physical origin of the frequency dependence is different. In the electrical lumped-parameter RC model the frequency dependence arises due to the finite time required to charge/discharge a capacitor shunted by a resistance. In the Drude model, the origin of the frequency dependence is due to electron scattering with impurities, defects, etc. having a characteristic exponential relaxation time constant (which can range from $\tau\sim$0.01-2 ps depending on the SLG mobility, Fermi level and velocity\cite{LinQE20,DasRMP83,TseAPL93}). The Drude model for SLG suggests that for low frequencies $\omega\ll 2\pi/\tau$ the real part of the conductivity is approximately identical to the DC conductivity, $\sigma_1 \sim\sigma_0$ with $\sigma_2\sim0$, whereas at higher frequencies $\sigma_1 \propto \omega^2$ and $\sigma_2 \propto \omega$. Identical dependencies also emerge for the case of a SLG channel represented as a \textit{RC} network with time constant $\tau'=C_G R_G$, as shown in the second part of Eqs.(\ref{eq:RA}, \ref{eq:XA}). However, in the latter case the frequency dependence arises due to $C_G$.

Fig.\ref{graph:Transport_1} shows typical S-parameter measurements of our Au Thru devices and the extracted transmission line parameters, resistance (R), inductance (L), conductance (B) and capacitance (C), using Eq.(\ref{eq:zed_matrix}). The transmission $(S_{21},S_{12})$ and reflection $(S_{11},S_{22})$ parameters demonstrate a small power dissipation $P_d = P_i (1-|S_{11}|^2-|S_{21}|^2)\sim0.16 mW$ in the Thru device up to 13.5GHz, with incident power $P_i=1mW$. This remains constant up to 7GHz, then increases slightly to$\sim0.17mW$ at higher frequencies, due to Eddy current losses in the metallic conductors\cite{AwanBOOK, DuffBOOK}. In contrast, the inductance and capacitance have frequency-independent response up to 13.5GHz, as expected, whereas the conductance has only a slight dependence, due to substrate losses.

Fig.\ref{graph:Transport_2} plots the measured scattering parameters, apparent resistance $R_A$ and reactance $X_A$, and their calculated values based on Eqs.(\ref{eq:RA},\ref{eq:XA}), for our GCPW at frequencies up to 13.5GHz at room temperature (at 0dBm and with $V_{BG}=0V$ in Fig.\ref{graph:Conductivity}). In contrast with the Au Thru device, the $S_{21}$ and $S_{12}$ parameters show RF and microwave transmission to be less than -20dB (or 100mV/V) at frequencies up to 13.5GHz, due to impedance mismatch with the measurement system, as expected based on the DC 2-probe resistance measurements. This level of transmission is found to be sufficient, combined with the calibration of the measurement system and employing parasitic impedance de-embedding, to extract the intrinsic electromagnetic properties of the SLG channel. A comparison of the power transmission, reflection and absorption coefficients of our GCPW with the Au Thru waveguide is shown in Table \ref{table1}. For $f<1 GHz$, $\sim88\%$ of the RF and microwave power is reflected, whilst 0.4\% is transmitted through the SLG channel and 12\% is absorbed (similar to the reference Thru device simulated and measured in Figs.\ref{graph:CPW},\ref{graph:Transport_1}). In contrast, at 13.5GHz the reflected power reduces to$\sim81\%$ and transmitted power increases to 0.9\%, whereas the absorbed power increases to$\sim18\%$.
\begin{figure*}
\centerline{\includegraphics[width=180mm]{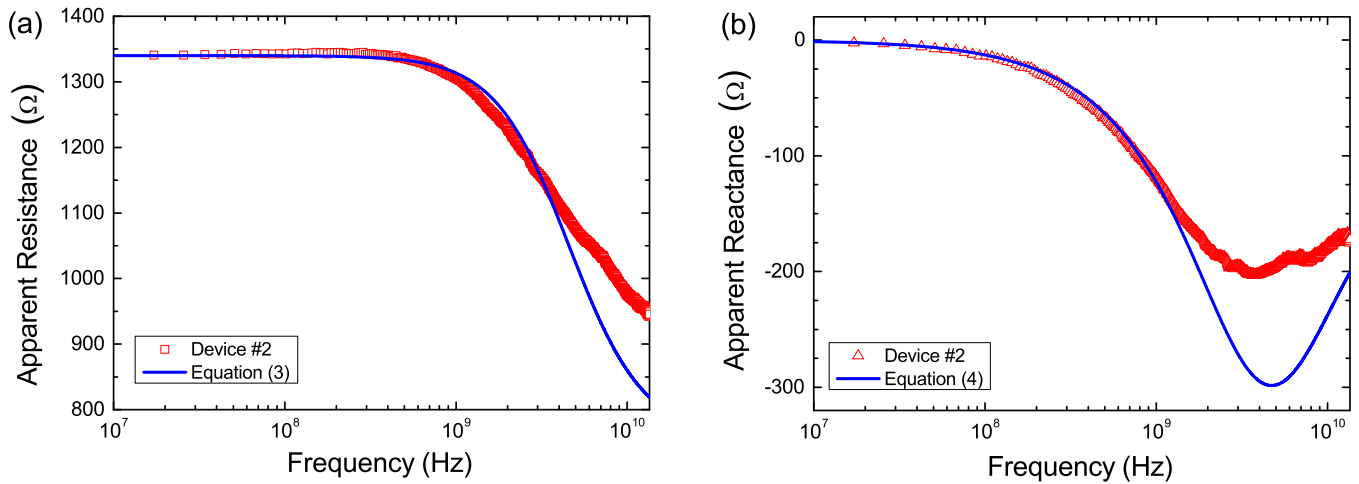}}
\caption{Comparison of measured (a) GCPW\#2 apparent resistance with Eqn.(\ref{eq:RA}), and (b) the corresponding measured apparent reactance with Eq.(\ref{eq:XA}) at frequencies up to 13.5GHz.}
\label{graph:device}
\end{figure*}
\begin{table}[h]\footnotesize
\begin{tabular}{|c|c|c|c|c|}
\hline
• & Au-Thru & Au-Thru & GCPW & GCPW \\
• & $f<1GHz$ & $f\sim13.5 GHz$ & $f<1GHz$ & $f\sim13.5 GHz$ \\
\hline
Transmission & 83.6\% & 82.6\% & 0.4\% & 0.9\% \\
\hline
Reflection & 0.7\% & 0.8\% & 87.6\% & 81.3\% \\
\hline
Absorption & 15.7\% & 16.6\% & 12\% & 17.8\% \\
\hline
\end{tabular}
\caption{RF and microwave power transmission, reflection and absorption in a Au-Thru CPW and GCPW.}
\label{table1}
\end{table}
\begin{table}[h]\footnotesize
\begin{tabular}{|c|c|c|c|c|c|c|}
\hline
• & $\sigma_0 (\Omega^{-1})$ & $R_G (\Omega)$ & $C_G(fF)$ & $R_C(\Omega)$ & $C_C (pF)$ & $\tau (ps)$ \\
\hline
GCPW \#1 & 28.9$e^2/h$ & 891 & 2.4 & 289 & 0.12 & 2.1 \\
\hline
GCPW \#2 & 34.1$e^2/h$ & 760 & 2.1 & 292 & 0.15 & 1.6 \\
\hline
\end{tabular}
\caption{GCPW parameters extracted from our DC and RF measurements at room temperature and 0V back-gate}
\label{table2}
\end{table}

A least-squares fit to the measured apparent resistance and reactance in Fig.\ref{graph:Transport_2} gives $R_C\sim289 \Omega$/contact and $C_C\sim0.12pF$/contact, with $C_{PS}=1.2 fF$, $C_{P0}=13.5 fF$ and $R_{P0}=50k\Omega$. For $f<1 GHz$ $R_A$ is constant with frequency, and its measured magnitude is $R_A\sim1.48 k\Omega$, in close agreement with the 2-probe DC value of $R_{2P}\sim1.47 k\Omega$. Thus, the intrinsic DC conductivity of our SLG channel is $\sigma _0\sim28.9 e^2/h$ after removal of the contact resistance. For $f>1 GHz$, $R_A$ rapidly decreases to $\approx 1 k\Omega$ at 13.5GHz, indicative of the presence of contact capacitance and resistance in the waveguide device (represented by the first part of Eq.(\ref{eq:RA})). We get $R_G \sim 0.89 \pm 0.014  k\Omega$ and $C_G\sim2.4\pm 0.065 fF$, with $R_G$ equivalent to a 4-probe resistance extracted from a 2-probe measurement. This extraction of 4-probe resistance from a 2-probe measurement, deploying our RF and microwave method, could also be useful for other applications where an independent method is needed for comparison with standard DC 4-probe measurements. Furthermore, when devices are inherently 2-probe, our RF and microwave method could extract their intrinsic 4-probe properties.

The measured $X_A$ up to 13.5GHz, given in Fig.\ref{graph:Transport_2}, shows good agreement with Eq.\ref{eq:XA}, to within $\pm$2.7$\%$ uncertainty for reactance measurements. The measured and calculated $X_A$ show a minima at$\sim$6GHz, due to the presence of $C_C$, as for Fig.\ref{graph:CPW}(b). The inset in Fig.\ref{graph:Transport_2} shows $X_A$ converted to an apparent capacitance. Fig.\ref{graph:device} plots the measured apparent resistance and reactance of a second GCPW device at 0dBm and 0V back-gate. We get $R_{2P}\sim1.34k\Omega$. Also shown for comparison is the calculated response based on the model of Fig.\ref{graph:CPW}  and Eqs.(\ref{eq:RA},\ref{eq:XA}). Reasonably good qualitative agreement is observed between the data in Figs.\ref{graph:Transport_2} and \ref{graph:device}. The corresponding extracted device parameters are given in Table II. The $C_G$ extracted for the two GCPW devices, $\sim2.4$ and 2.1fF, are compatible to the estimate\cite{YuPNAS110} using $C_G=enab/(V_{BG}-V_D)\sim11.6fF$. The quantum capacitance\cite{YuPNAS110} is found to be $C_Q\sim e^{2}D\sim4.7pF$, where the density of states $D=gabE_F/(\pi\hbar^{2}v_F^{2})$ and $g=4$ is the spin and valley degeneracy. The total capacitance is the series combination of the geometric and quantum capacitance given by $C_T=1/(1/C_G+1/C_Q)\sim11.6fF$ and is close to the geometric capacitance, since $C_G\ll C_Q$.

The time constants for the two devices are $\tau=C_G R_G \sim2.1$ and $\sim1.6ps$, respectively, (which compare well with $\tau\sim1.1ps$ reported in Ref.[\onlinecite{LiN4}]). The former leads to $R_G '\sim R_G (1-0.032)$ at 13.5GHz (from the second part in Eq.(\ref{eq:RA})). However, $R_G$ and the corresponding real components of the conductivity $\sigma_1 \sim 29 e^2/h$ and $\sigma_1\sim 34 e^2/h$ for the two GCPW devices, are found to be frequency-independent up to 13.5GHz, since $R_G$ is the same for the entire 0.01 to 13.5GHz range. Thus, the intrinsic electromagnetic response of SLG when carrying RF and microwave signals does not involve any additional power dissipation with respect to that at DC (i.e. Joule heating) up to 13.5GHz within our experimental uncertainty of $\pm$1.6$\%$. These results also suggest that the background of absorption in the single-particle optical gap $\hbar \omega < 2 |E_F|$ is very small, in agreement with recent reports \cite{WoesNM14}. This background, not captured by the single-particle (Drude) model considered here,  is due to electron-electron interactions \cite{PrinPRB88} and vanishes  at increasing $\omega$ following a $\omega^{-1}$ relation \cite{PrinPRB88}.

Static and dynamic conductivities are found to be identical, i.e. $\sigma_0\sim\sigma_1$ up to 13.5GHz. This represents a unique property of graphene in comparison with other materials, such as metals\cite{Poza, DuffBOOK} and superconductors\cite{AwanIPSMT149}, when carrying transport RF and microwave signals. This may have significant impact on the future design of ultra high-speed electronic devices based on graphene (potentially up to $\sim$THz based on extrapolation of the models and their agreement with our results reported here, albeit to 13.5GHz), as well as their eventual very large scale integration (VLSI) into integrated circuits, particularly in comparison with the high dissipation ($\sim$100Wcm$^{-2}$) typically found in current state-of-the-art interconnects and Si based processors operating at few GHz\cite{BakerRJ}.
\section{\label{Conclusion}Conclusions}
We reported the room-temperature transport conductivity of graphene at frequencies up to 13.5GHz measured using coplanar waveguides. This is independent of the frequency of the applied transport current in the DC to microwave range, so that the real and imaginary components of the complex AC dynamic conductivity are $\sigma_1\sim\sigma_0$  and $\sigma_2 \sim0$, suggesting negligible additional power dissipation at high-frequencies compared to DC. Our results are in good agreement with the Drude model for dynamic conductivity with a momentum relaxation time $\tau\sim2.1ps$ and $\sim1.6ps$, for the two measured devices. This contrasts the quadratic frequency-dependence usually found in metals and superconductors and is promising for the future potential applications of graphene in ultra high-speed electronic devices. In particular, the measured frequency-independent resistance suggests it may be possible to realize broadband and low noise radiation detectors, high-frequency low noise amplifiers and mixers. We also anticipate potential applications in a variety of radio-frequency and microwave sensors enabling spectroscopic detection of physical or biological properties of materials and substances (i.e. in the case of biosensors antibody-antigen interactions at the graphene surface), in contrast with existing static conductance change based sensors.
\section{\label{Acknol}Acknowledgements}
We acknowledge funding from EU Graphene Flagship (no. 604391), ERC Grant Hetero2D, EPSRC Grants EP/K01711X/1, EP/K017144/1, RR/105758, Wolfson College, a Royal Society Wolfson Research Merit Award and M. Polini, A. Awan, B. Kibble, I. Robinson, N. Ridler, G. Pan, P. Davey, M. Z. Ahmed, L. Garcia-Gancedo, A. Katsounaros, M. Luukkainen, S. Ellil\"{a}, G. Fisher, S. Wordingham and C. Barnett for useful discussions.

\end{document}